\documentclass[iop,showkeys]{emulateapj}

\submitted{Accepted April 2, 2013, for publication in the Astrophysical Journal Letters
}
\begin{document}

\title{Molybdenum, Ruthenium, and the Heavy r-process Elements in Moderately Metal-Poor Main-Sequence Turnoff Stars}
\author{Ruth C. Peterson$^1$}
\affil{$^1$Astrophysical Advances}

\begin{abstract}

The ratios of elemental abundances observed in metal-poor stars of the Galactic
halo provide a unique present-day record of the
nucleosynthesis products of its earliest stars. While the heaviest elements
were synthesized by the r- and s-processes,
dominant production mechanisms of
light trans-ironic elements were obscure until recently.
This work investigates further our 2011 conclusion
that the low-entropy regime of a high-entropy wind (HEW)
produced molybdenum and ruthenium in two moderately metal-poor turnoff stars
that showed extreme overabundances of those elements with respect to iron.
Only a few, rare nucleosynthesis events may have been involved.
	
Here we determine abundances for Mo, Ru, and other trans-Fe elements
for 28 similar stars by matching spectral calculations to well-exposed
near-UV Keck HIRES spectra obtained for beryllium abundances.
In each of the 26 turnoff stars with Mo or Ru line detections
and no evidence for s-process production (therefore old),
we find Mo and Ru to be three to six times overabundant. In contrast,
the maximum overabundance is reduced to factors of three and two for
the neighboring elements zirconium and palladium.
Since the overproduction peaks sharply at Mo and Ru,
a low-entropy HEW is confirmed as its origin.
	
The overabundance level of the heavy r-process elements varies
significantly, from none to a factor of four, but is uncorrelated with Mo and Ru
overabundances. Despite their moderate metallicity, stars in this group
trace the products of
different nucleosynthetic events: possibly very few events, possibly
events whose output depended on environment, metallicity, or time.
	
\end{abstract}

\section{Introduction}

The heavy-element abundance distributions of metal-poor stars, reviewed by \citet{2008ARA&A..46..241S},
can yield critical diagnostics of the objects and environments that formed the material,
and its incorporation into Galactic halo and disk stars. For the heaviest elements, these processes are reasonably well understood. In single stars of metallicity below one-thirtieth solar, [Fe/H] $<$ --1.5, elements
from barium ($Z$ = 56) onward are produced by rapid neutron addition on iron-peak seed nuclei in the $r$-process. Their elemental abundance ratios are preserved, even though their overall level with respect to iron can be more than an order of magnitude greater or less than the solar level. Only in more metal-rich single stars
do elements begin to appear that are created by the $s$-process (slow neutron
capture), in pulsations in intermediate-mass asymptotic giant branch (AGB) stars.
The AGB evolutionary time of a few 100 Myr suggests a time delay of this order
in the formation of such stars.

In contrast, many processes are invoked for the trans-Fe elements gallium 
through cadmium ($Z$ = 31 to 48). \citet{2011ApJ...742...21P} provides
a summary. \citet{2012A&A...545A..31H} emphasize that multiple processes are required to explain light trans-Fe abundance ratios, especially at the lowest metallicities. \citet{2013A&A...550A.122S} find that electron-capture (O-Ne-Mg) supernovae may contribute the lightest trans-Fe elements, especially 
if $r$-process content is extremely low. For molydenum and ruthenium (Mo, Ru; $Z$ = 42, 44) in stars with moderate $r$-process levels, recent work favors the low-entropy
domain of a high-entropy wind (HEW) above the neutron star formed
in a Type II supernova \citep[e.g.][]{1999ApJ...516..381F}. 

In the solar system, 
\citet{2009PASA...26..194F} reproduced all seven of the solar isotopes of
molybdenum by selecting models from a parameterized grid of HEW calculations.
They find it ``can co-produce the light $p$-, $s$-, and $r$-process isotopes
between Zn ($Z$ = 30) and Ru ($Z$ = 44) at ... 
low entropies $S$ $\leq$ 100 -- 150. Under these
conditions, the light trans-Fe elements are produced in a charged-particle
($\alpha$-) process, including all $p$-nuclei up to $^{96,98}$Ru. 
... This nucleosynthesis component is primary.''

Support for HEW production in metal-poor stars with low $r$-process content 
emerged from observed ratios of the light trans-Fe element 
yttrium and the heavy $r$-process element europium (Y, Eu; $Z$ = 39, 63). 
\citet{2007A&A...476..935F} noted an 
anti-correlation between [Y/Eu] and [Eu/Fe], and \citet{2010ApJ...724..975R} reproduced 
the [Y/Eu] ratios via HEW models \citep{2010ApJ...712.1359F}.

Strong confirmation for HEW production in metal-poor stars emerged when 
\citet{2011ApJ...742...21P} determined Mo and Ru abundances 
from ${\rm Mo\,{\textsc{ii}}}$ and ${\rm Ru\,{\textsc{ii}}}$ lines 
in ultraviolet 
spectra of five unevolved stars.
Two of these, HD 94028 and HD 160617 with [Fe/H] = $-$1.4 and $-$1.8, showed 
[Mo/Fe] = +1.0 and +0.8, and [Ru/Fe] = +0.7 and +0.6.     
Zr ($Z$ = 40) was less enhanced, as were the  $r$- and $s$-process heavy elements 
Only HEW models have 
predicted high excesses of the light trans-Fe elements 
that are confined to this narrow mass range in $Z$. 

\citet{2011ApJ...742...21P} noted that existing ${\rm Mo\,{\textsc{i}}}$ abundances for $>$20 field and cluster giants with similarly low heavy $r$-process content, [Eu/Fe] $<$ +0.6, all show [Mo/Fe] $<$ +0.5. This rarity of high Mo excesses implied 
that only a few distinct nucleosynthesis events produced the light trans-Fe elements in the two extreme turnoff stars. 

In this work, we derive abundances for Mo, Ru, and other trans-Fe elements 
in 28 additional moderately 
metal-poor turnoff stars. We discuss results for 26 of these, 
excluding one (HD 106038) that shows mild $s$-process contamination, and one (G 66-30) 
whose high temperature and broader lines suggest it is a blue straggler. 
We derive Mo and Ru abundances from the ${\rm Mo\,{\textsc{i}}}$ line at 3864.10\AA\ and the ${\rm Ru\,{\textsc{i}}}$ line at 3498.94\AA, 
lines which were often used in previous analyses of giants.
A few stars have only upper limits or marginal detections; for the rest, the 
two elements exhibit the same excess to $\pm$0.1 dex. In all cases, the mean MoRu 
abundance is enhanced by a factor of three to six above the solar proportion: +0.4 $\leq$ [MoRu/Fe] $\leq$ +0.8. 

We also derive and discuss the abundances of 
the lighter trans-Fe elements strontium, yttrium, and zirconium (Sr, Y, and Zr; $Z$ = 38, 39, 40), plus the heavier element palladium (Pd; $Z$ = 46), the latter analyzed 
to date in fourteen stars. None of these elements shows enhancements as large 
as that of Mo and Ru in any star. From this we confirm the low-entropy regime 
of a HEW as the principal means of production of Mo and Ru in moderately 
metal-poor turnoff stars. 

We measure the overall $r$-process not from europium, but from dysprosium and 
erbium (Eu, Dy, Er; $Z$ = 63, 66, 68), 
because the lines of ${\rm Dy\,{\textsc{ii}}}$ and ${\rm Er\,{\textsc{ii}}}$ are less blended. 
We find that 
the $r$-process enhancement varies from none to a factor of four, 
varying significantly from star to star, implying distinct events.

\section{Stellar Spectra} 

\begin{deluxetable*}{rccccrrrrrrrrrr}
\tabletypesize{\scriptsize}
\tablecaption{Stellar Parameters and Element Abundances [Element/Fe]\label{tbl-st}}
\tablehead{
\colhead{Star} 
& \colhead{$T_{{\rm eff}}$}
& \colhead{log {\it g}}
& \colhead{[Fe/H]}
& \colhead{$V_{{\rm t}}$}
& \colhead{Mn}
& \colhead{Co}
& \colhead{Sr}
& \colhead{Y}
& \colhead{Zr}
& \colhead{Mo}
& \colhead{Ru}
& \colhead{Pd}
& \colhead{Nd}
& \colhead{Eu}
} 
\startdata                                                            
BD -8 4501  &  6100  &  4.2  & $-$1.50  &  1.1  & $-$0.2  &  0.0  & $-$0.2  &  0.0  &  0.3  &  0.7  &  0.7  & 0.2 &  0.1  &  0.5 \\
                                                            
BD -17 484  &  6300  &  4.2  & $-$1.50  &  1.1  & $-$0.2  &  0.0  & $-$0.1  &  0.1  &  0.3  &  $<$0.5  &  0.5  &    &  0.0  &  0.3 \\
                                                            
BD +4 4551  &  6000  &  4.1  & $-$1.30  &  1.0  & $-$0.3  & $-$0.1  &    0.0  &  0.2  &  0.4  &  0.6  &  0.6  & 0.2 &  0.0  &  0.2 \\
                                                            
BD +7 4841  &  6000  &  3.9  & $-$1.50  &  1.1  & $-$0.3  &  0.0  &  0.0  &  0.1  &  0.4  &  0.6  &  0.6  &    &  0.0  &  0.2 \\
                                                            
BD +13 3683  &  6000  &  4.0  & $-$1.85  &  1.2  & $-$0.2  &  0.1  &    0.0  &  0.1  &  0.4  &  $\leq$0.8  &  0.8  &    &  0.0  &  0.5 \\
                                                            
BD +17 4708  &  6150  &  3.9  & $-$1.70  &  1.1  & $-$0.1  &  0.1  &  0.0  &  0.0  &  0.3  &  $<$0.6  &  $\leq$0.6  &    &  0.0  &  0.4 \\
                                                            
BD +21 607  &  6150  &  4.1  & $-$1.70  &  1.1  & $-$0.2  &  0.1  & $-$0.2  &  0.0  &  0.2  &  $<$0.6  &  0.6  &    &  0.0  &  0.3 \\
                                                            
BD +23 3912  &  5800  &  3.6  & $-$1.45  &  1.2  & $-$0.2  &  0.1  & $-$0.1  &  0.1  &  0.3  &  0.5  &  0.5  &  0.2  & $-$0.2  &  0.1 \\
                                                            
BD +36 2165  &  6200  &  4.1  & $-$1.60  &  1.1  & $-$0.1  &  0.0  & $-$0.1  &  0.0  &  0.3  &  $<$0.5  &  0.5  &    &  0.0  &  0.4 \\
                                                            
BD +37 1458  &  5550  &  3.6  & $-$1.90  &  1.2  & $-$0.3  &  0.0  &  0.0  &  0.3  &  0.5  &  0.7  &  0.7  &    &  0.1  &  0.5 \\
                                                            
BD +42 3607  &  5900  &  4.4  & $-$2.10  &  1.1  & $-$0.3  &  0.2  &    0.0  &  0.0  &  0.4  &  $<$0.6  &  $\leq$0.6  &    &  0.0  &  0.3 \\
                                                            
BD +51 1696  &  5600  &  4.5  & $-$1.30  &  1.0  & $-$0.2  &  0.0  & $-$0.3  & $-$0.1  &  0.2  &  0.5  &  0.5  &  0.3  &  0.0  &  0.5 \\
                                                            
G 113-9  &  6200  &  4.2  & $-$1.60  &  1.1  & $-$0.4  &  0.0  &  0.0  &  0.2  &  0.4  &  0.7  &  0.7  & 0.0 &  0.0  &  0.0 \\
                                                            
G 115-49  &  5900  &  4.4  & $-$2.10  &  1.0  & $-$0.2  &  0.3  & $-$0.3  & $-$0.1  &  0.3  &  $<$0.6  &  $\leq$0.6  &    &  0.0  &  0.2 \\
                                                            
G 180-24  &  6050  &  4.1  & $-$1.50  &  1.1  & $-$0.2  &  0.1  &  0.0  &  0.1  &  0.4  &  0.6  &  0.6  & 0.3 &  0.0  &  0.3 \\
                                                            
G 188-22  &  6000  &  4.1  & $-$1.30  &  1.1  & $-$0.3  & $-$0.1  &    0.0  &  0.2  &  0.4  &  0.6  &  0.6  &    &  0.0  &  0.2 \\
                                                            
G 191-55  &  6000  &  4.3  & $-$1.75  &  1.1  & $-$0.4  &  0.0  & $-$0.1  &  0.0  &  0.3  &  0.7  &  0.7  & 0.3 &  0.0  &  0.4 \\
                                                            
G 192-43  &  6200  &  3.9  & $-$1.50  &  1.2  & $-$0.1  &  0.0  & $-$0.2  &  0.1  &  0.3  &  0.6  &  0.6  & 0.2 & $-$0.1  &  0.6 \\
                                                            
G 66-30  &  6400  &  4.1  & $-$1.50  &  1.2  & $-$0.1  &  0.0  &    0.0  &  0.0  &  0.2  &  $<$0.9  &  $<$0.9  &    &  0.0  &  0.5 \\
                                                            
HD  31128  &  5950  &  4.2  & $-$1.50  &  1.0  & $-$0.2  &  0.0  &    0.0  &  0.1  &  0.3  &  0.6  &  0.6  &    & $-$0.1  &  0.2 \\
                                                            
HD 106038  &  6100  &  4.2  & $-$1.30  &  1.1  & $-$0.2  &  0.1  &  0.1  &  0.5  &  0.6  &  0.8  &  0.8  &    &  0.3  &  0.3 \\
                                                            
HD 108177  &  6100  &  4.1  & $-$1.75  &  1.1  & $-$0.1  &  0.1  & $-$0.1  &  0.0  &  0.3  &  $<$0.6  &  0.6  &    &  0.0  &  0.2 \\
                                                            
HD 160617  &  6000  &  3.8  & $-$1.80  &  1.2  &  0.0  &  0.0  &  0.0  &  0.0  &  0.4  &  0.8  &  0.6  &    &  0.3  &  0.4 \\
                                                            
HD 161770  &  5650  &  3.6  & $-$1.60  &  1.2  & $-$0.3  &  0.0  & $-$0.1  &  0.0  &  0.2  &  0.5  &  0.5  &  0.2  &  0.0  &  0.3 \\
                                                            
HD 188510  &  5450  &  4.55  & $-$1.55  &  0.9  & $-$0.1  &  0.0  & $-$0.1  & $-$0.1  &  0.2  &  0.4  &  0.4  &  0.2  &  0.0  &  0.3 \\
                                                            
HD 194598  &  5900  &  4.0  & $-$1.20  &  1.1  & $-$0.1  &  0.0  & $-$0.3  & $-$0.1  &  0.2  &  0.4  &  0.4  &  0.1  &  0.0  &  0.3 \\
                                                            
HD 233511  &  6100  &  4.3  & $-$1.55  &  1.1  & $-$0.3  &  0.1  & $-$0.1  &  0.1  &  0.4  &  0.5  &  0.5  & 0.1 &  0.0  &  0.0 \\
                                                            
HD 241253  &  5750  &  4.0  & $-$1.30  &  1.0  & $-$0.1  &  0.1  &  0.0  &  0.1  &  0.2  &  0.6  &  0.6  &  0.2  &  0.0  &  0.3 \\
                                                            
HD 247168  &  5700  &  4.3  & $-$1.60  &  1.0  & $-$0.2  & $-$0.1  & $-$0.4  & $-$0.2  &  0.0  &  0.5  &  0.5  &  0.0  &  0.0  &  0.3 \\

\enddata
\hskip 40pt \tablecomments{Units: $T_{{\rm eff}}$, $\mathrm{K}$; $V_{{\rm t}}$, {km~s\ensuremath{^{-1}}}.
Marginal detections and non-detections are indicated by $\leq$ and $<$ respectively.
}
\end{deluxetable*}

Our data are a subset of the Keck HIRES echelle spectra that 
\citet[Table 1]{2011ApJ...743..140B} obtained to measure beryllium in over a hundred metal-poor stars 
near the main-sequence turnoff. We selected spectra of stars with 
metallicities $-$2.0 $\leq$ [Fe/H] $\leq$ $-$1.4 
from the 16 runs that followed the Keck CCD upgrade, from September 2004 to July 2010. Currently we 
have analyzed only the 3440\AA\ -- 3950\AA\ portion of the bluest CCD, except for fourteen stars where 
reductions reached the ${\rm Pd\,{\textsc{i}}}$ line at 3404.579\AA.

We reduced all spectra from the raw images in the IRAF\footnote 
{IRAF is distributed by the National Optical Astronomy Observatories, which are operated 
by the Association of Universities for Research in Astronomy, Inc., under cooperative 
agreement with the National Science Foundation.} environment. 
We performed bias and dark removal, 
coadded (with cosmic-ray removal) any multiple spectral images of the same object obtained the same night, extracted orders with removal of sky and local interorder background, 
corrected the dispersion using concurrent Th-Ar exposures, 
rectified the continuum with spline fits, 
and spliced together adjacent orders. The spectra 
were shifted to zero velocity by cross-correlation against theoretical templates, using 
the routine \emph {fxcor}. 

For the strongest-lined stars, no true continuum could be discerned throughout the blue region. After a preliminary analysis, we re-rectified the continuum by first 
dividing the extracted echelle spectrum 
order-by-order by the best-fit theoretical spectrum, itself normalized by dividing 
by the continuum spectrum included in the calculation (after fitting by hand 
a pseudocontinuum to the 3860\AA\ -- 3900\AA\ Balmer-limit region). 
We then ran spline fits on the ratio, and divided 
the extracted echelle orders by the fits. This largely but not entirely removes 
the continuum suppression due to atomic- and molecular-line blending, provided the 
the theoretical spectrum is a close match. 

\section{Synthetic Spectral Analysis}

Stellar parameters and abundances were derived by matching each stellar spectral observation to theoretical spectra calculated for each star using an updated version of the \citet{1993KurCD..18.....K} SYNTHE program with the stellar models of \citet{2003IAUS..210P.A20C}. We input a list of molecular and atomic 
line transitions with wavelengths, energy levels, and gf-values, and a model atmosphere 
characterized by effective temperature $T_{{\rm eff}}$, surface gravity log {\it g}, 
microturbulent velocity $V_{{\rm t}}$, and logarithmic iron-to-hydrogen ratio [Fe/H] 
with respect to that of the Sun. We calculate the entire spectral region, and compare against each observed spectrum to find the best match. \citet[Section 4]{2011ApJ...742...21P} provides details.

In deriving the stellar parameters, rather than use photometry, we match strengths line-by-line. For this range of line strengths, the 3463.1\AA\ -- 3469.2\AA\ region was most useful, 
with additional ${\rm Fe\,{\textsc{ii}}}$ lines and 
lower-excitation ${\rm Fe\,{\textsc{i}}}$ lines from 3440\AA\ -- 3550\AA, and the very strong low-excitation 
${\rm Fe\,{\textsc{i}}}$ lines at 3820\AA\ -- 3830\AA. Except for the latter, we used weak lines only.

We started with the \citet[Table 2]{2011ApJ...743..140B} values for $T_{{\rm eff}}$, log $g$, and 
[Fe/H]. We checked [Fe/H] from ${\rm Fe\,{\textsc{i}}}$ lines with lower excitation potential 
$\sim$ 3 eV, then compared ${\rm Fe\,{\textsc{i}}}$ lines of 0.8 -- 1.5 eV versus high-excitation ${\rm Fe\,{\textsc{i}}}$ lines to set $T_{{\rm eff}}$. We then confirmed or altered log $g$ 
from the match to ${\rm Fe\,{\textsc{ii}}}$ lines and the wings of the strong ${\rm Fe\,{\textsc{i}}}$ lines. As these log $g$ indicators always agreed to 0.1\,dex and yielded reasonable values, non-LTE effects are evidently small. \citet{2001ApJ...559..372P} note that this consistency among $T_{{\rm eff}}$ and log $g$ indicators in metal-poor turnoff stars extends to visible wavelengths and to Balmer lines.  It requires the extensive line-list modifications described there, and the use of atmospheric models with no convective overshoot, including our adopted \citet{2003IAUS..210P.A20C} models. Where available, the near-UV flux level and slope are matched as well, 
suggesting the overall temperature scale is reliable. 

We 
simply adopted microturbulent velocities based on stellar parameters, choosing the 
solar value of 1.0 km/s for main-sequence stars of near-solar $T_{{\rm eff}}$, 0.9 km/s for 
cooler main-sequence stars, and 1.1 to 1.2 km/s for progressively hotter and/or 
lower gravity stars, as suggested from the convection discussion 
of \citet{1997A&A...318..841C}. 

\begin{figure*}[ht!]
\epsscale{1.00}
\plotone{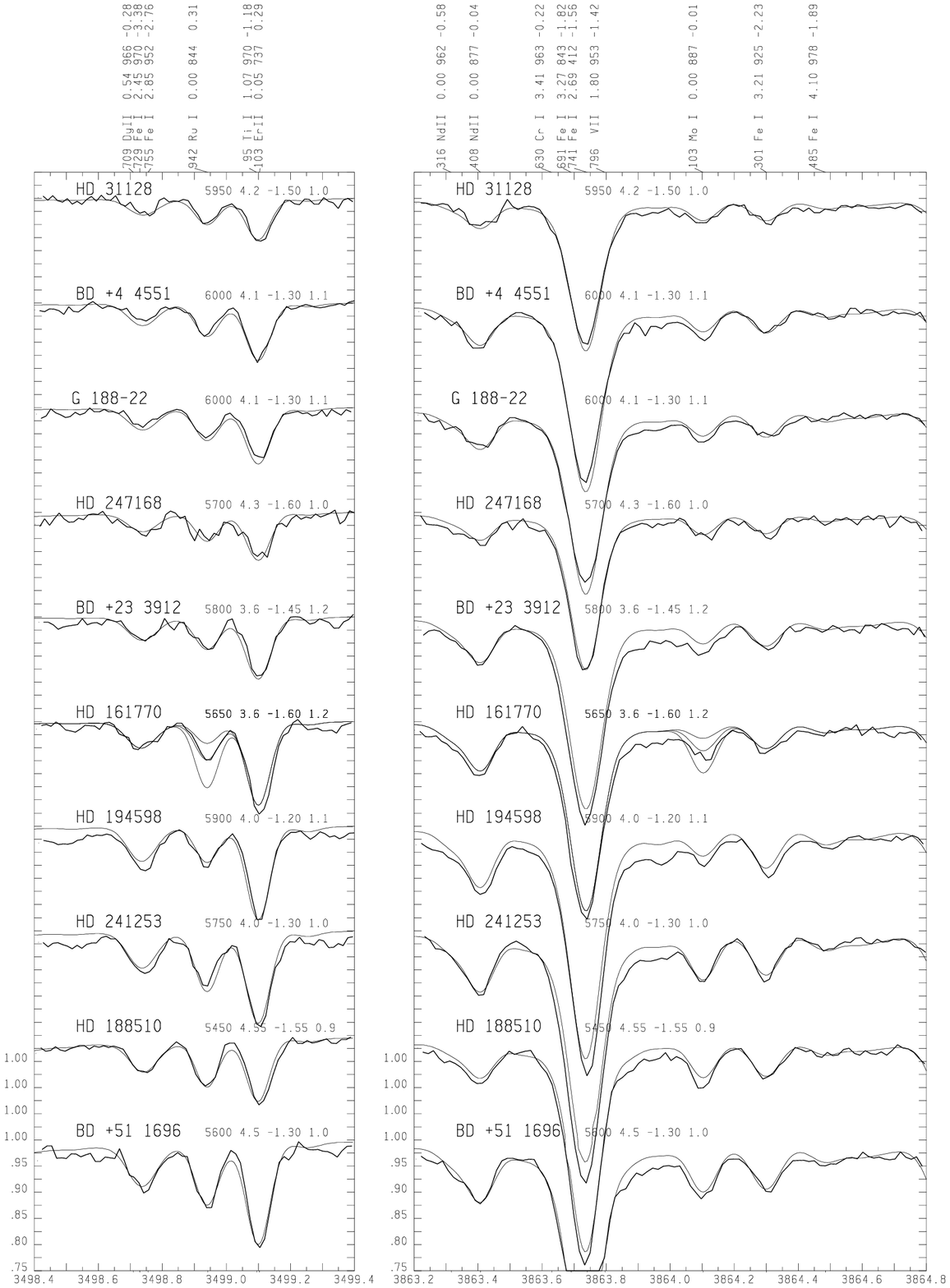}
\caption{
Comparisons are shown between observed and calculated spectra
in two spectral regions, with 
wavelength in \AA ngstroms given at the bottom.
Plots for ten stronger-lined stars are offset vertically by
20\% of the normalized continuum;
ticks on the y axis indicate 2.5\% of this level.
The heavy line is the observed spectrum, and the light line
the calculated spectrum. Its strongest lines 
are identified at the top. First are the digits following
the decimal place of the line center wavelength, 
then its species, its lower excitation potential in eV,
its strength (stronger lines have lower numbers), and its log $g$f-value.
The star identification from Table~\ref{tbl-st} is given above each plot.
Following it are the stellar parameters adopted for its calculation: 
$T_{{\rm eff}}$, log {\it g}, [Fe/H], and $V_{{\rm t}}$. The 
relative elemental abundances adopted are provided in Table~\ref{tbl-st}. 
For HD 161770, calculations are also shown with Mo/Ru abundances 0.3 dex
higher and lower. 
~
}
\label{fig:fig-moru1hi}
\end{figure*}

Table~\ref{tbl-st} lists the resulting stellar model parameters and abundances. 
By repeating selected stellar parameter
determinations from different starting points, we estimate uncertainties of $\pm$100\,K
in $T_{{\rm eff}}$, $\pm$0.2\,dex in log $g$, and $\pm$0.1\,dex in [Fe/H]. 
Due to the low excitation and ionization potentials of
the ${\rm Mo\,{\textsc{i}}}$ and ${\rm Ru\,{\textsc{i}}}$ lines, [Mo/Fe] and [Ru/Fe] values rise by  $\sim$0.05\,dex for a 100\,K rise in 
$T_{{\rm eff}}$.

The uncertainty estimates are supported by the agreement of our values with those 
\citet{2011ApJ...743..140B} found for the same stars. Excluding BD +13 3683, 
mean differences and 1$\sigma$ mutual standard deviations of $T_{{\rm eff}}$, log $g$, 
and [Fe/H] are 30\,K, 0.19\,dex, and 0.02\,dex, and 154\,K, 0.28\,dex, and 0.15\,dex. 
However,  
our temperatures are occasionally $>$200\,K hotter (G 115-49, BD +13 3683, BD +42 3607) or cooler (BD -8 4501, BD +51 1696, HD 241253).

BD +13 3683 is the most deviant star. 
Our values are higher for $T_{{\rm eff}}$ by 500\,K, for 
log $g$ by 0.9 dex, and for [Fe/H] by 0.5 dex.
The near-UV spectrum of this star is 
a close match to that of HD 108177, which has the same parameters as those of 
BD +13 3683 to within the uncertainties.
We infer that BD +13 3683 is a binary, a turnoff primary with a cooler companion, 
and that our reliance on the 3500\AA\ region
has led to a higher $T_{{\rm eff}}$.

\section{Molybdenum, Ruthenium, and Palladium Abundances}

\begin{figure*}[ht!]
\epsscale{1.00}
\plotone{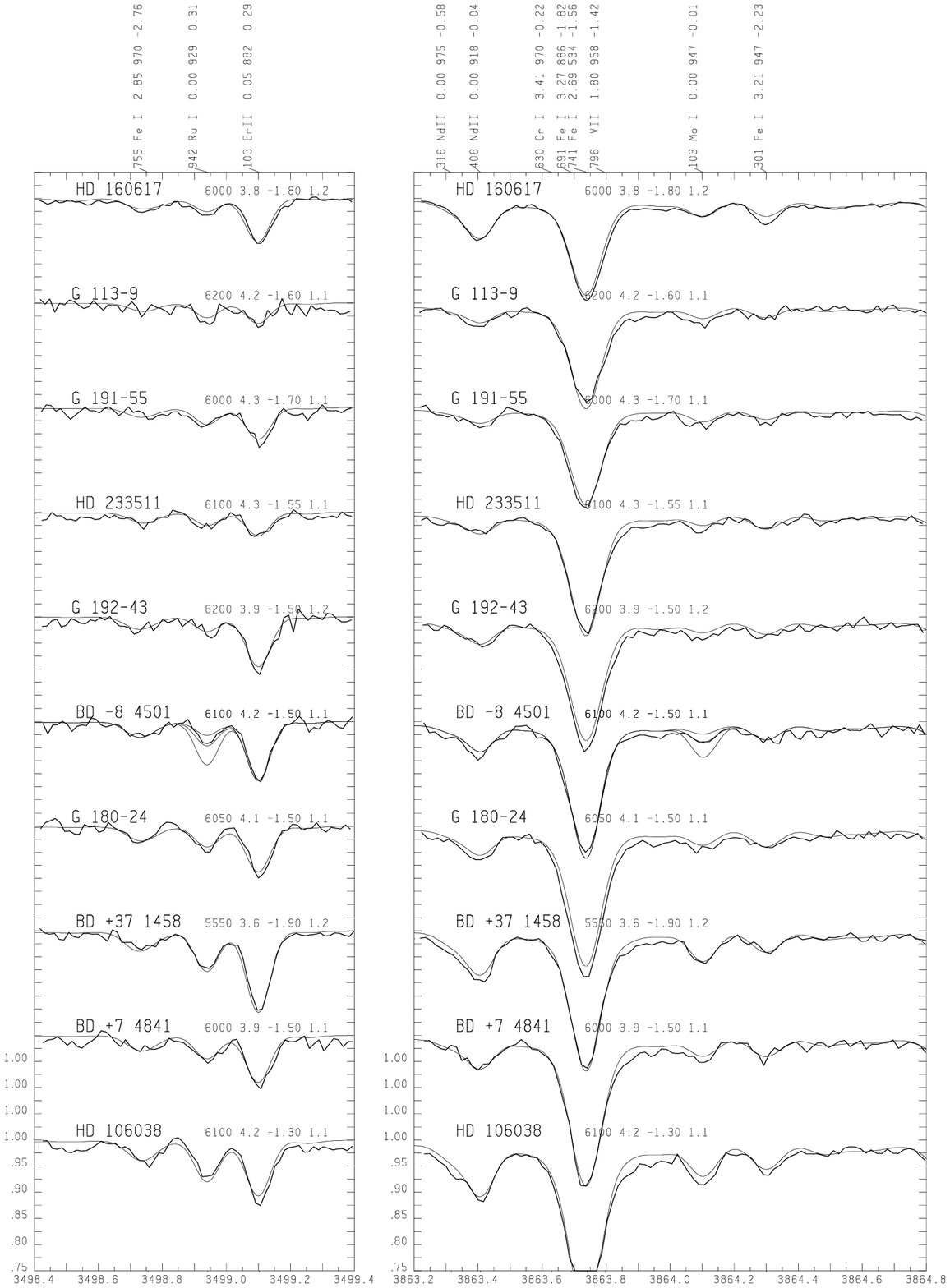}
\caption{
The same comparisons as in Figure~\ref{fig:fig-moru1hi} are shown for
ten weaker-lined stars. Nine are newly analyzed in this work. The top star,
HD 160617, analyzed by \citet{2011ApJ...742...21P}, is shown for comparison. 
As for HD 161770 in Figure~\ref{fig:fig-moru1hi},
calculations adopting Mo/Ru abundances 0.3 dex
higher and lower are shown for BD -8 4501.
}
\label{fig:fig-moru2hi}
\end{figure*}

For simplicity in the plots below and for better accuracy in the comparison of 
the Mo and Ru abundances versus those of lighter and heavier elements, we have
tabulated the average excess derived from the Mo and Ru lines, instead of individual 
values for each element. The spectral comparisons presented immediately below 
indicate that the two lines
give abundances that agree to within 0.1 dex whenever both lines are detected. 

Figures~\ref{fig:fig-moru1hi} and \ref{fig:fig-moru2hi} compare the calculations 
based on the parameters in Table~\ref{tbl-st}
to the observations around the ${\rm Mo\,{\textsc{i}}}$ line at 3864.103\AA\ and the ${\rm Ru\,{\textsc{i}}}$ line at 3498.942\AA. These regions also show the $r$-process line of ${\rm Er\,{\textsc{ii}}}$ at 3499.103\AA, and the $s$-process lines of ${\rm Nd\,{\textsc{ii}}}$ near 3863.4\AA. 

Figure~\ref{fig:fig-moru1hi} plots the stronger-lined stars, and Figure~\ref{fig:fig-moru2hi} shows weaker-lined ones. 
Figure~\ref{fig:fig-moru2hi} also includes HD 160617 for reference, comparing a calculation using the updated line list used throughout this work 
to the UVES pipeline spectrum analyzed by \citet{2011ApJ...742...21P}. 
For one star in each figure -- HD 161770 in Figure~\ref{fig:fig-moru1hi} and BD -8 4501 in Figure~\ref{fig:fig-moru2hi} -- calculations are also shown in which [Mo/Fe] and [Ru/Fe] values higher and lower by 0.3 dex are adopted. Since the lines are weak and minimally blended, this doubles and halves the line strength. 

From this 
we estimate a measurement uncertainty of $\pm$0.1\,dex in all cases where both lines 
are detected. The uncertainty is $\pm$0.15\,dex for four stars where ${\rm Ru\,{\textsc{i}}}$ is detected 
but ${\rm Mo\,{\textsc{i}}}$ is not: BD $-$17 484, BD +21 607, BD +36 2165, and HD 108177.
It is $\pm$0.2\,dex for the three stars where even ${\rm Ru\,{\textsc{i}}}$ 
is only marginally detected: BD +17 4708, BD +42 3607, and G 115-49. We nonetheless 
include these stars throughout this discussion.

The mean [MoRu/Fe] value for the 26 stars is 0.58 $\pm$ 0.02\,dex, 
an average enhancement of a factor of four above solar.
Several stars have enhancements similar to those of the two extreme stars of 
\citet{2011ApJ...742...21P}, HD 160617 and HD 94028. 
Consequently, no longer is the incorporation of products from very few nucleosynthesis events necessarily implied for these two stars.

We have not detected an intrinsic spread 
in the average abundance of Mo and Ru among these 26 stars. The 1$\sigma$ standard 
deviation of the individual [MoRu/Fe] 
values is 0.09 dex, which is attributable to the uncertainties alone. 
A trend is hinted with metallicity, but it relies heavily on the least certain [MoRu/Fe] values. 

\begin{figure*}[ht!]
\epsscale{1.10}
\plotone{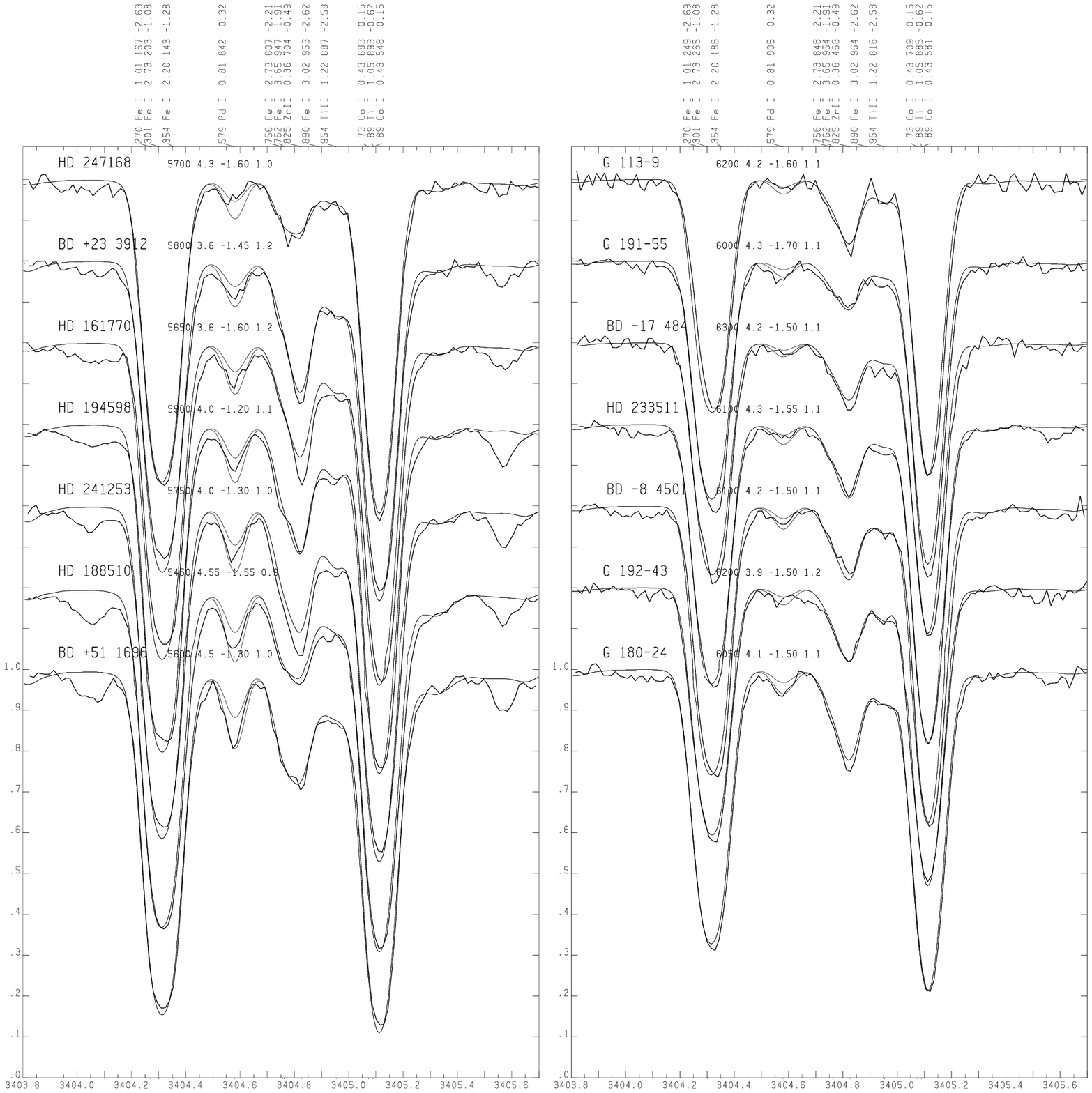}
\caption{
Similar comparisons for seven strong-lined and seven weak-lined stars are shown for the spectral region containing a ${\rm Pd\,{\textsc{i}}}$ line at 3404.579\AA. The 
calculations show two choices for palladium, [Pd/Fe] = 
+0.0 (solar) and +0.3 (twice solar). The [Pd/Fe] values listed in Table~\ref{tbl-st} are derived from this plot. They range over 0.0 $\leq$ [Pd/Fe] $\leq$ +0.3; values as high as those of [Mo/Fe] and [Ru/Fe] are ruled out.
}
\label{fig:fig-moru3hi}
\end{figure*}

In fourteen stars, we also
analyzed the ${\rm Pd\,{\textsc{i}}}$ line at 3404.579\AA.
Fits to that spectral region are shown in
Figure~\ref{fig:fig-moru3hi}. Strong-lined stars are in the left panel, and
weaker-lined stars in the right. For each star, two abundances for Pd were adopted, [Pd/Fe] = 0.0 and +0.3, a factor-of-two increase.
All the observed ${\rm Pd\,{\textsc{i}}}$ lines fall on or
within these two values. Consequently, in these stars Pd 
is always substantially less enhanced than Mo and Ru.

\section{Heavy $s$- and $r$-process Abundances}

For the $s$-process indicator we used not lanthanum but 
neodymium (La, Nd; $Z$ = 57, 60), 
based on the ${\rm Nd\,{\textsc{ii}}}$ lines near 3863\AA, as the
strong ${\rm La\,{\textsc{ii}}}$ lines of  \citet{2011ApJ...742...21P}
lie redward.
We adopted a solar value of log(Nd/(H+He)) = $-$10.54.

As seen from Table 1 and directly 
in Figures 1 and 2, the ${\rm Nd\,{\textsc{ii}}}$ lines are detected in all stars. 
All but three show the solar proportion or higher, [Nd/Fe] $\geq$ 0.0. 
Only for [Eu/Fe] $\geq$ +0.3 is this expected from an $r$-process contribution alone. 
For HD 106038, the result [Nd/Fe] = +0.3 is attributed to the $s$-process, because 
[Eu/Fe] = +0.3 but [Zr/Fe] = +0.6, the highest value of all the 29 turnoff 
stars. HD 160617, analyzed previously, 
is the only other star in Table 1 showing [Nd/Fe] $>$ 0.2.

The values of [Sr/Fe], [Y/Fe], and [Zr/Fe] track one another closely. 
[Sr/Zr] and [Y/Zr] average $-$0.41 $\pm$ 0.02\,dex and $-$0.26 $\pm$ 0.01\,dex, with 
1$\sigma$ deviations of 0.08\,dex and 0.06\,dex. Any trend for [Zr/Fe] to 
follow [MoRu/Fe] is weak. We do see an anti-correlation 
between [Y/Eu] and [Eu/Fe]: [Y/Eu] = -1.17 $\times$ [Eu/Fe] + 0.10, 
with a scatter of about 0.1 dex about this relationship.  
For HD 106038, [Y/Eu] is 0.45\,dex higher than the trend; we attribute 
this to its $s$-process enhancement.

For the $r$-process indicator we used the ${\rm Dy\,{\textsc{ii}}}$ lines at 3531.707\AA\ and 3536.019\AA\ 
plus the ${\rm Er\,{\textsc{ii}}}$ line at 3499.103\AA. 
We estimate an uncertainty of 0.08\,dex in this measurement, based on the comparison 
of the ${\rm Dy\,{\textsc{ii}}}$ and ${\rm Er\,{\textsc{ii}}}$ line fits. Figures 
1 and 2 confirm that the ${\rm Er\,{\textsc{ii}}}$ line is easily detected and 
extremely well matched.

The ${\rm Eu\,{\textsc{ii}}}$ line at 3819.7\AA\ provided a 
secondary check only, as it is blended by ${\rm Fe\,{\textsc{i}}}$ at 3819.493\AA\ and ${\rm Cr\,{\textsc{i}}}$ at 
3819.565\AA, and falls on the deep wing of a nearby Balmer line. 
Our spectral calculations for $r$-process elements are run assuming abundances derived 
from the \citet{1999ApJ...525..886A} $r$-process fractions. 
We thus express the Dy/Er $r$-process abundance excesses 
in terms of [Eu/Fe], for consistency with other work.

We find a mean [Eu/Fe] = 0.29 $\pm$ 0.03\,dex. The average $r$-process enhancement 
for these 26 stars is then a factor of two higher than solar, and a factor of two 
lower than the enhancement of Mo and Ru. No correlation is present between the 
two. 

[Eu/Fe] does show a significant intrinsic spread. 
The observed 1$\sigma$ spread of 0.14\,dex and the 0.08 uncertainty imply
an intrinsic dispersion of 0.12\,dex. Moreover, 
in Figure~2, HD 233511 and BD -8 4501 have substantially 
different ${\rm Er\,{\textsc{ii}}}$ line strengths, despite having virtually the 
same stellar parameters.

\section{Implications for Nucleosynthesis}

This work establishes that moderate to high excesses of molybdenum and ruthenium are common among mildly metal-poor stars. Mo and Ru are enhanced similarly, 
by an average factor of four, but Zr and Pd 
are always less overabundant. This substantiates HEW as the source in metal-poor stars of the light trans-Fe elements with $Z$ $\sim$ 44, as only the low-entropy regime of HEW predicts the sizable overproduction of just these elements.

The lower [Mo/Fe] values previously obtained for giants, using the same 
${\rm Mo\,{\textsc{i}}}$ line, remain puzzling. 
Non-LTE effects or model uncertainties may be worse in giant analysis, 
as the cooler giant models are more transparent and more susceptible 
to the effects of convection. Illustrative is the \citet{2012ApJS..203...27R} analysis of UV and optical spectra of four metal-poor subgiants and giants, yielding $T_{{\rm eff}}$ values $>$ 200\,K cooler than estimates from $V - K$ colors, and ${\rm Fe\,{\textsc{i}}}$ abundances that varied systematically with wavelength by up to 0.3\,dex. Our turnoff analyses show no such effects (Sect.\ 3). 

The difference 
might equally well result from a dependence of low-entropy HEW production on metallicity, since most of the previous analyses are for stars of lower metallicity  than these; or on the field halo versus globular-cluster environment, since many previously analyzed giants are members of globular clusters.

Because high molybdenum and ruthenium abundances are typical 
of moderately metal-poor turnoff stars, 
exceptionally few nucleosynthesis events are not required for the high values 
\citet{2011ApJ...742...21P} found for HD 94028 and HD 160617. 
However, the group as a whole does show a star-to-star 
spread in [Eu/Fe]. Either these 
stars typically did incorporate limited and diverse subsets 
of the ensemble of nucleosynthetic events, or that ensemble itself depended on 
local environment, metallicity, or time. 

In any case, the oldest halo turnoff stars of roughly one-thirtieth solar metallicity do indeed provide tracers of a range of nucleosynthetic events. The majority of these favored production of Mo and Ru more heavily than did nucleosynthesis at later times. With a larger sample of unevolved stars 
that span a larger metallicity range, the detailed abundance distributions can be 
correlated against space motions and metallicities to try to identify the defining 
characteristics of the progenitor stars, or of the subsequent incorporation 
of their products, that has led to this nucleosynthetic diversity.

\acknowledgements 
We thank M. Spite and W. Aoki for helpful discussions, and the anonymous referee and a second referee (Chris Sneden) 
for comments that significantly improved this paper.
Critical to this work are observations made with 
the Keck Observatory HIRES spectrograph, under programs H03aH, H11aH, H30aH, H41aH, H169Hb, H177Hb, and H233Hb (PI A. Boesgaard). We appreciate their expert efforts. The previously-analyzed observation of HD 160617
was made with the ESO Telescope and UVES spectrograph at the Paranal Observatory, under program 065.L-0507(A).
This research has made use of the Keck Observatory Archive (KOA), which is 
operated by the W. M. Keck Observatory and the NASA Exoplanet Science Institute 
(NExScI), under contract with the National Aeronautics and Space Administration.

\bibliography{sdmorur}

\end{document}